# Development of holmium-163 electron-capture spectroscopy with transition-edge sensors

**M. P. Croce • M. W. Rabin • V. Mocko • G. J. Kunde • E. R. Birnbaum • E. M. Bond • J. W. Engle • A. S. Hoover • F. M. Nortier • A. D. Pollington • W. A. Taylor • N. R. Weisse-Bernstein • L. E. Wolfsberg**

*Los Alamos National Laboratory, Los Alamos, NM, USA*

**J. P. Hays-Wehle • D. R. Schmidt • D. S. Swetz • J. N. Ullom**

*National Institute of Standards and Technology, Boulder, CO, USA*

**T. E. Barnhart • R. J. Nickles**

*University of Wisconsin, Madison, WI, USA*

**Abstract** Calorimetric decay energy spectroscopy of electron-capture-decaying isotopes is a promising method to achieve the sensitivity required for electron neutrino mass measurement. The very low total nuclear decay energy ($Q_{EC}$ < 3 keV) and short half-life (4570 y) of $^{163}$Ho make it attractive for high-precision electron capture spectroscopy (ECS) near the kinematic endpoint, where the neutrino momentum goes to zero. In the ECS approach, an electron-capture-decaying isotope is embedded inside a microcalorimeter designed to capture and measure the energy of all the decay radiation except that of the escaping neutrino. We have developed a complete process for proton-irradiation-based isotope production, isolation, and purification of $^{163}$Ho. We have developed transition-edge sensors for this measurement and methods for incorporating $^{163}$Ho into high-resolution microcalorimeters, and have measured the electron-capture spectrum of $^{163}$Ho. We present our work in these areas and discuss the measured spectrum and its comparison to current theory.

## 1 Introduction

The mass of the neutrino has wide-ranging implications, significantly influencing the evolution of large-scale cosmic structure, big-bang nucleosynthesis, and our understanding of basic symmetries and asymmetries of the universe. Strong experimental evidence of neutrino flavor oscillations implies that the neutrino has a small, nonzero mass. The best results to date suggest that the neutrino mass scale is 0.05-1 eV, about a million times smaller than the mass of the electron, making the neutrino the lightest of all the massive particles and making mass measurement extremely difficult [1, 2].

Calorimetric decay energy spectroscopy of electron-capture decaying isotopes is one measurement method that promises to achieve the required mass sensitivity [3-8]. In this approach, an electron-capture-decaying isotope is embedded completely inside a cryogenic microcalorimeter designed to capture and measure the energy of all the decay radiation except that of the escaping neutrino. The very low total nuclear decay energy ($Q_{EC}$ < 3 keV), 100% decay by electron capture, and short half-life (4570 years) of $^{163}$Ho make this rare, unusual, synthetic isotope suited for this measurement challenge. The central challenges for this approach are isotope production and purification; incorporation of $^{163}$Ho into sensors; high resolution spectroscopy of electron-capture decays; scaling up to tens of thousands of sensors; and understanding the nuclear and atomic physics of this decay to determine the neutrino mass from the kinematic endpoint. Here we present our work to date along these lines.

## 2 Production and isolation of $^{163}$Ho

$^{163}$Ho does not occur naturally, and is not commercially available. We explored a variety of production options, described in Ref. [9]. Proton irradiation of isotopically natural Dy has some advantages over other production routes, including relatively high isotopic purity of $^{163}$Ho and ease of obtaining the isotopically natural Dy target material. We have performed three proton irradiations of natural Dy, summarized in Table 1. The first target ("Foil 1") was a foil of natural Dy obtained from Alfa Aesar. It was irradiated for 10 hours in a PETtrace cyclotron with an average proton current of 10 $\mu$A and proton energy at the target of 16.1±0.1 MeV [10]. The irradiated Foil 1 is shown in Fig. 1 (left). Discoloration from heating is clearly visible, and indicates the beam position on the target. Foil 1 contained 377 mg of natural Dy and 94 $\mu$g of naturally occurring $^{165}$Ho (250 ppm). Based on mass spectrometry of the F1 product solution, radiochemical yield, and a geometric factor for the beam profile, we estimate that the total mass of $^{163}$Ho in the irradiated Foil 1 was 60 ng. For a second irradiation we wanted to reduce the amount of unnecessary stable $^{165}$Ho in the separated product solutions, so we obtained a Dy foil ("Foil 2") of much higher chemical purity from Ames Laboratory. Foil 2 contained only 2.8±1 ppm of natural $^{165}$Ho, a factor of approximately 90 better than Foil 1. Foil 2 (Fig. 1, center) was also irradiated with the PETtrace cyclotron with the same beam conditions as Foil 1, but for 40 hours.

These two irradiated foils provided sufficient $^{163}$Ho for initial experiments. A third, much larger target was irradiated at the Los Alamos National Laboratory Isotope Production Facility (IPF 1) to supply $^{163}$Ho for a future large-scale experiment with high neutrino mass sensitivity. This target consists of 12.6 g of high chemical purity, isotopically natural Dy in an Inconel case. It was irradiated at an average proton beam current of 230 $\mu$A with nominal incident energy 25 MeV and nominal exit energy 10 MeV, and a total cumulative charge of 33,645 $\mu$Ah. The irradiated target IPF 1 is shown in Fig. 1 (right). It is estimated to contain approximately 145 $\mu$g of $^{163}$Ho corresponding to 2.48 MBq. A second, similar target (IPF 2) is available for a future irradiation. Table 1 summarizes properties of the Dy targets, irradiation conditions and estimated quantity of produced $^{163}$Ho.

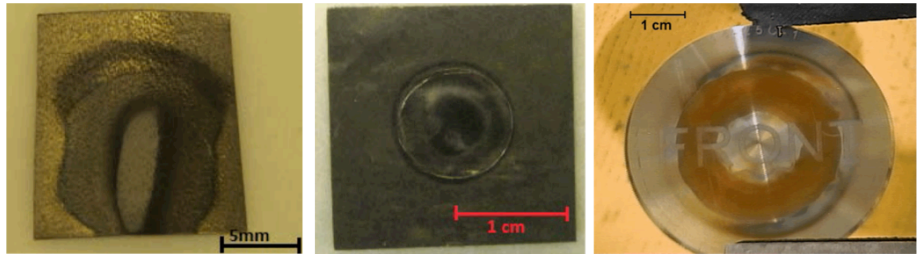

**Fig. 1** Three proton irradiated natural Dy targets: Foil 1 (left), Foil 2 (center), and IPF 1 target in Inconel casing (right). (Color figure online)

Holmium from portions of Foil 1 and Foil 2 were isolated by cation-exchange high performance liquid chromatography (HPLC) by a procedure described in Refs. [10, 11]. Table 2 summarizes the content of residual Dy, stable $^{165}$Ho, and $^{163}$Ho in isolated product solutions. The entire 54.5 mg center section of Foil 1, where the majority of $^{163}$Ho was located, was used for product solution F1. First, the HPLC separation procedure was optimized to achieve high separation resolution of Ho from Dy using natural Ho and Dy solutions. Following this, Ho from the center section of Foil 1 was separated using a 2.5 x 25 cm column filled with AG50W-X8 resin and 70 mM α-HIBA, pH 4.6 eluent. The collected Ho fraction was processed in two steps by ion-exchange chromatography and extraction chromatography to remove the α-HIBA and concentrate the Ho. The product solution was measured by mass spectrometry to have 24.8±1.3 ng $^{163}$Ho, 401±15 ng Dy, and a $^{165}$Ho/$^{163}$Ho ratio of 397±15. Overall recovery of Ho from the center section of Foil 1 was 72±3%, determined by measuring the concentration of $^{165}$Ho in the initial foil and the concentration of $^{165}$Ho in the F1 product solution. A 12 mg wedge-shaped portion of the center of Foil 2 was used for Ho-Dy separation with a similar procedure to obtain product solution F2C1 (Fig. 2). Relative to product solution F1, an improved Ho-Dy separation factor was obtained for F2C1 by using a smaller portion of the Dy target (12 mg instead of 54.5 mg) and a lower eluent pH of 4.46 instead of 4.6. The chromatogram of natural Ho-Dy separation using these conditions is shown in Fig. 2, left. The Foil 2 product solution contained 13.2±0.4 ng $^{163}$Ho, 40.5±1.3 ng Dy, and a $^{165}$Ho/$^{163}$Ho ratio of 1.8±0.1, determined by mass spectrometry.

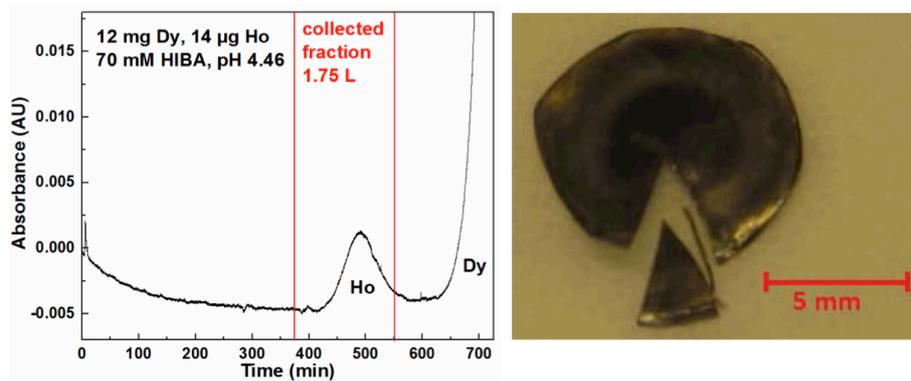

**Fig. 2** (Left) Chromatogram of separation of natural Ho-Dy to identify Ho fraction to be collected from Foil 2. (Right) Center of Foil 2 with wedge-shaped piece used for $^{163}$Ho product solution F2C1. (Color figure online)

| Target ID | Proton irradiation conditions | Dy mass | Initial $^{165}$Ho mass | Estimated $^{163}$Ho mass |
|---|---|---|---|---|
| Foil 1 | 100 $\mu$Ah, 16.1±0.1 MeV | 377 mg | 94 $\mu$g | 60 ng |
| Foil 2 | 400 $\mu$Ah, 16.1±0.1 MeV | 677 mg | 1.9 $\mu$g | 170 ng |
| IPF 1 | 33,645 $\mu$Ah, 10-25 MeV | 12.6 g | 35.3 $\mu$g | 145 $\mu$g |

**Table 1** Summary of natural Dy targets irradiated for $^{163}$Ho production

| Product ID | Source | Residual Dy mass | $^{165}$Ho/$^{163}$Ho mass ratio | $^{163}$Ho mass |
|---|---|---|---|---|
| F1 | Entire center of Foil 1, 54.5 mg | 401±15 ng | 397±15 | 24.8±1.3 ng |
| F2C1 | Wedge of Foil 2, 12 mg | 40.5±1.3 ng | 1.8±0.1 | 13.2±0.4 ng |

**Table 2** Summary of Ho and Dy content, measured by mass spectrometry, in isolated $^{163}$Ho product solutions after irradiation, separation, and purification.

### 3 Detector development

The unique requirements of $^{163}$Ho electron capture spectroscopy motivated the development of a new style of transition-edge sensor (TES). The primary concerns were low heat capacity so that the dynamic range of the sensor could be matched to the low Q value of $^{163}$Ho decay, and a mechanically robust design that would be compatible with a wide range of absorber attachment options during the prototyping phase of the project [12]. The TESs for this experiment consist of a 350 $\mu$m square Mo-Cu bilayer with superconducting transition temperature near 110 mK, located on a 50 $\mu$m wide silicon beam for thermal isolation (Fig. 3). The silicon beam is the full wafer thickness of 275 $\mu$m, and is defined by a deep reactive ion etch process. A pad at the end of the beam provides an absorber attachment point. Heat capacity of the TES and beam structure is approximately 0.9 pJ/K at 100 mK.

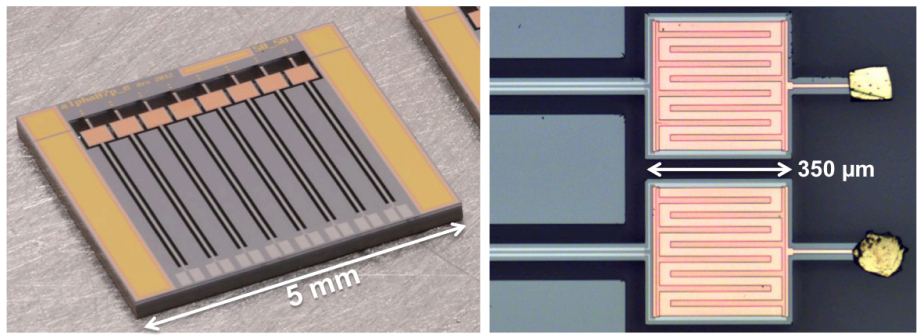

**Fig. 3** Transition-edge sensors developed for $^{163}$Ho electron capture spectroscopy use a silicon beam thermal isolation to enable varied absorber attachments. Each die contains 8 independent TESs. Gold absorbers with embedded electron-capture-decaying isotopes are attached to the end of the beam. (Color figure online)

During the first phase of the project, while $^{163}$Ho was being produced, testing of these transition-edge sensors began with $^{55}$Fe ($t_{1/2}$=2.74 y, 100% EC). This isotope was easily available, and provides energy peaks at 6539 eV, 769 eV, and 82 eV that correspond to the electron binding energies of the Mn daughter atom (Fig. 4, left). This allowed characterization of the detector response in the energy range of interest. A solution of $^{55}$Fe, where $^{55}$Fe represented 15% of the total iron mass on May 23, 2012, was prepared at Los Alamos National Laboratory. This solution contained a much lower fraction of stable isotopes than commercially available $^{55}$Fe solutions. The $^{55}$Fe was electroplated onto a gold foil with activity of approximately 500 Bq in a 1 mm$^2$ area. Small portions of this gold foil with electroplated $^{55}$Fe were cut with a typical size of approximately 50x100x15 $\mu$m, folded in half, and diffusion welded on a hot plate in a nitrogen atmosphere at 400°C for 1 hour to encapsulate the radioactive material. The absorbers were attached to the TESs with either a low-temperature (150°C) diffusion bond or an indium bump bond. Both absorber attachment methods provided similar detector performance, but the indium bump bond method proved to be more reliable. The best energy resolution from this type of detector, given by the Gaussian component of a Bortels function fit, was 7.5±0.2 eV FWHM at 6539 eV with a tail factor of 10 eV (Fig. 4, right) [13].

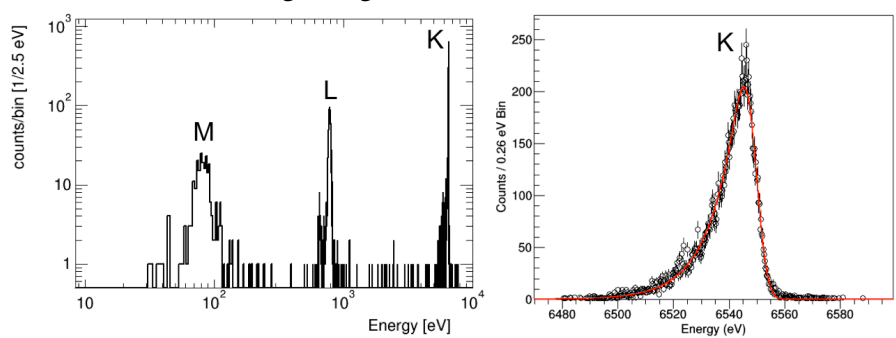

**Fig. 4** Measurements with encapsulated $^{55}$Fe enabled detector development while $^{163}$Ho was being produced. (Left) Energy peaks corresponding to the K, L, and M electron binding energies of the Mn daughter atom span the energy range of interest. (Right) The best measured energy resolution was 7.5±0.2 eV FWHM at 6539 eV. (Color figure online)

Experiments with $^{55}$Fe, as well as our other work on Q spectroscopy of embedded actinides [14], suggested several important characteristics of an optimal absorber with embedded radioactive material. First, radioisotopic and chemical purity of the embedded material is critical. Additional material in the deposit, besides the radioactive isotope of interest, adds deleterious heat capacity and increases the potential for energy trapping in the deposit. Second, the radioactive material in the absorber matrix must be distributed in grains that are small compared to the relevant radiation path lengths in order to provide a uniform environment for energy deposition and thermalization.

**4 Incorporation of $^{163}$Ho into sensors**
Initial experiments with $^{163}$Ho product F1 demonstrated the presence of excess stable impurities, in particular a $^{165}$Ho/$^{163}$Ho ratio of 397±15 (see Table 2), that led to extremely poor energy resolution (Fig. 5, left). This spectrum resulted from a deposit of the Foil 1 product in 0.1 M HCl solution, dried on a gold foil. The gold foil was folded and pressed several times in an

**Development of Holmium-163 electron capture spectroscopy…**

attempt to homogenize the deposit within the gold matrix and improve the energy thermalization environment as in [14], then folded and pressed several more times with a small piece of $^{55}$Fe electroplated on gold foil to provide peaks for energy calibration. Pulses from $^{55}$Fe were observed to have a different shape than those from $^{163}$Ho, interpreted to be the result of the different chemical and physical forms of each isotope within the absorber. This allowed removal of the $^{55}$Fe events from the spectrum in Fig. 5 (left), after they were used for energy calibration. Product solution F2C1 was much more pure with a $^{165}$Ho/$^{163}$Ho ratio of 1.8±0.1 as a result of less initial $^{165}$Ho in the Dy target material, and longer irradiation. A similarly prepared absorber, made by drying F2C1 product in 0.1M HCl solution on a gold foil, then folding and pressing the foil to encapsulate the deposit, resulted in the energy spectrum in Fig. 5 (right). The higher purity of the F2C1 product led to greatly improved energy resolution. With clearly defined peaks resulting from $^{163}$Ho decay, this spectrum was energy calibrated using the Dy daughter M complex and no $^{55}$Fe was added. This product solution was used for subsequent work to optimize the absorber structure and composition to further improve the measured energy resolution.

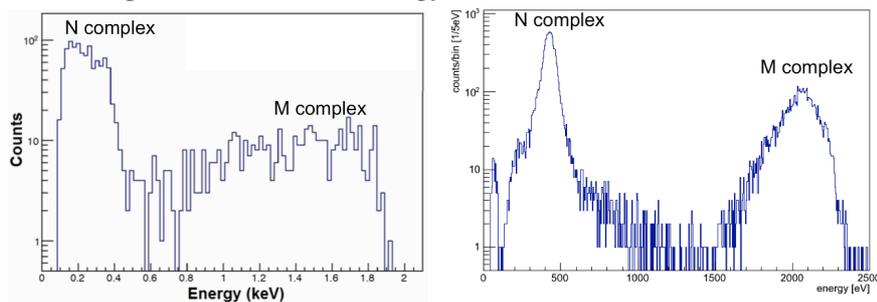

**Fig. 5** Foil 1 $^{163}$Ho product, with a $^{165}$Ho/$^{163}$Ho ratio of 397 +/- 15, resulted in the left spectrum with very poor energy resolution. A similarly prepared absorber with Foil 2 product, with a $^{165}$Ho/$^{163}$Ho ratio of 1.8 +/- 0.1, resulted in greatly improved energy resolution (right). (Color figure online)

We have focused on two methods to improve absorber structure for electron-capture spectroscopy of $^{163}$Ho: aqueous solution deposition in nanoporous gold (Fig. 6) to evenly distribute the $^{163}$Ho in small domains within the absorber, and annealing in forming gas to remove impurities. Fig. 7 shows the significant improvements in energy resolution resulting from these two methods. Nanoporous gold was produced using a method similar to that described in [15]. A 1.5 $\mu$m thick layer of Au-Ag alloy was electron-beam evaporated onto a 5 $\mu$m thick Au foil. The Ag was dealloyed by placing the foil in room-temperature 69% nitric acid for a period of 26 hours. The resulting foil had a nanoporous gold layer with pore sizes in the 50-100 nm range (Fig. 6). The nanoporous gold layer readily absorbs aqueous solutions. When these solutions dry, the crystals in the deposit are constrained by the pore size to form a nanocomposite structure. Occasionally, we have observed the formation of larger crystals on the surface of the nanoporous gold as well, which lead to reduced energy resolution. Because of the foil backing, small sections could be cut to form absorbers of desired heat capacity. A 30 $\mu$m diameter glass capillary was used to deposit F2C1 $^{163}$Ho product in 0.1M HCl solution into the nanoporous gold layer of ~50x100 $\mu$m sections of the foil. The foils were folded in half and pressed to encapsulate the deposits, then attached to transition-edge sensors.

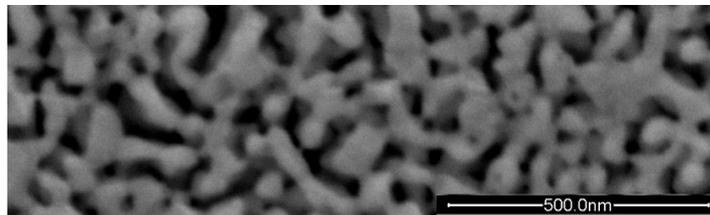

**Fig. 6** Nanoporous gold was used to constrain the crystal size in dried deposits of $^{163}$Ho product solution and create nanocomposite absorbers with embedded $^{163}$Ho.

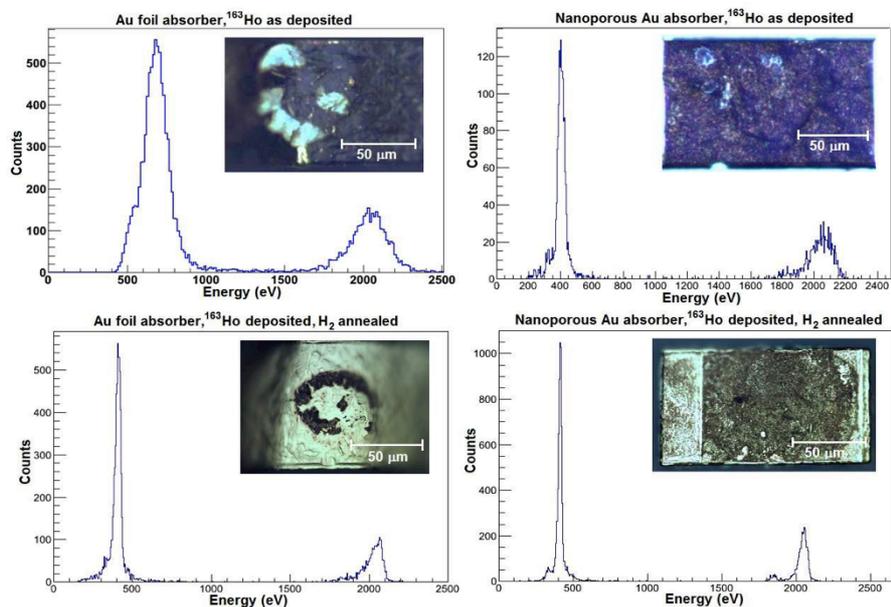

**Fig. 7** Improving the structure and composition of $^{163}$Ho deposits (insets) within the absorber led to significant improvements in measured energy resolution. (Color figure online)

**Development of Holmium-163 electron capture spectroscopy…**

The composition of the deposits was characterized by energy-dispersive x-ray spectroscopy (EDS) in a scanning electron microscope, which indicated the presence of significant amounts of Na, Cl, and organic compounds (introduced during separation and purification of $^{163}$Ho). We have demonstrated that the majority of these contaminants can be eliminated from the deposit by heating to 800°C for 6 hours in a 5% $H_2$, 95% $N_2$ atmosphere. EDS spectra of samples after this annealing process no longer show measurable Na or Cl. Absorbers made by heating the foil-backed nanoporous gold after deposition show greatly improved energy resolution (Fig. 7). The highest-resolution spectrum from an absorber made by drying a deposit of F2C1 $^{163}$Ho product solution in nanoporous gold, then hydrogen annealing to remove impurities, is shown in Fig. 8.

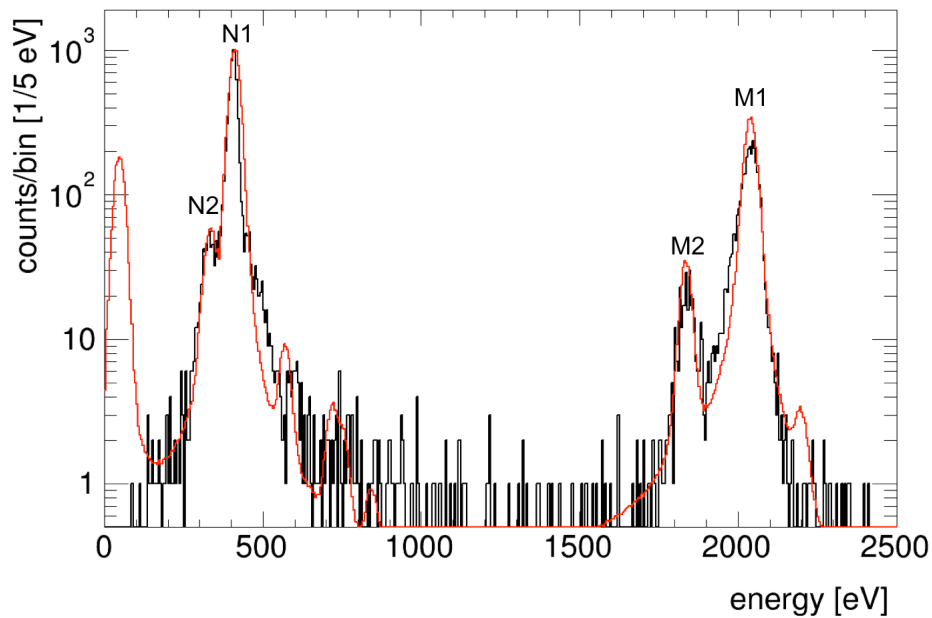

**Fig. 8** Measured electron-capture spectrum of $^{163}$Ho (histogram) with calculation overlaid (red curve) from Faessler et al. [18] convolved with a Gaussian distribution (35 eV FWHM) to represent detector reponse. (Color figure online)

## 5 Results

Fig. 8 shows our highest-resolution $^{163}$Ho electron-capture spectrum to date, corresponding to a 40 hour measurement with an activity of approximately 0.1 Bq. Caution is required outside of the strongest peaks in the spectrum. The statistics are still too weak to warrant a detailed quantitative discussion.

There are four major peaks in the spectrum, each tentatively labeled with a corresponding single-hole intermediate state of the $^{163}$Dy daughter atom. The highest energy peak is associated with a hole in the Dy M1 ($3s_{1/2}$; n=3, l=0, j=1/2) state with electron binding energy of 2040 eV, and is used for a single-point energy calibration. We associate the three remaining peaks with the respective Dy single-hole states and electron binding energies: M2 ($3p_{1/2}$; n=3, l=1, j=1/2) at 1836 eV, N1 ($4s_{1/2}$; n=4, l=0, j=1/2) at 411 eV, and N2 ($4p_{1/2}$; n=4, l=1, j=1/2) at 333 eV. The degree of alignment between measured peak locations and tabulated electron binding energies is strong, all within 1%, and independent of relative peak heights, widths, or spectral shape. This provides confidence in the assignment of each of the four major peaks to the corresponding $^{163}$Dy single-hole intermediate state. The O1 ($5s_{1/2}$; n=5, l=0, j=1/2) peak expected at 48 eV is below the trigger threshold.

We determined the spectral resolution by separately fitting the N and M spectral regions to a single-tail Bortels function, the convolution of a Gaussian (of width $\Delta E = 2.35\sigma$) with an exponential (1/e tailing factor $\tau$). This provides an empirically plausible fit for ECS, Q spectroscopy, and alpha spectroscopy. For the N spectral region, we find $\Delta E = 22$ eV and $\tau = 12$ eV. The M spectral region has $\Delta E = 43$ eV and $\tau = 28$ eV. Very similar sensors containing only $^{55}$Fe have shown $\Delta E = 7.5$ eV and $\tau = 10$ eV (Fig. 4). These values include the contribution of the natural line widths. For $^{163}$Ho ECS spectra, the larger $\Delta E$ relative to $^{55}$Fe and the dependence of resolution on energy suggest that the resolution is currently limited by factors not included in simple microcalorimeter models. We hypothesize this to be variability in the energy thermalization process within the absorber.

The theoretical spectral shape is an incoherent sum of modified Lorentzian (or Breit Wigner) peaks [3-5, 16-20] The modifications account for cutoff due to energy conservation and for the finite mass of the neutrino. Each peak corresponds to an excited electronic configuration of the $^{163}$Dy daughter atom, described using single-electron-type Dy orbitals. This requires specifying the deviations from the Dy electronic ground state for both holes and electrons, e.g. a vacancy in a normally filled M1 ($3s_{1/2}$; n=3, l=0, j=1/2) orbital and an electron in a normally vacant N7 ($4f_{7/2}$; n=4, l=3, j=7/2) orbital. The original treatment by De Rujula entails complicated excited daughter atom states, but focused on single-hole Dy states [3-5]. Robertson's skeptical assessment considered the role of more complex excitations (e.g. two-hole Dy states) and their potential negative impact on determining neutrino mass [20]. Faessler's detailed treatment addresses these issues with a relativistic calculation for the overlap and exchange corrections of 1-hole, 2-hole, and 3-hole states [16-18]. This last treatment appears the most complete, and we use it for comparison with our data.

Fig. 8 includes a calculated spectrum (red curve) based on Ref. [18], with a Q value of 2.8 keV [6, 21], vertically scaled so that the N1 peak height matches the data and broadened by convolution with a Gaussian of 35 eV FWHM. The use of a Gaussian broadening with no tailing is a simplification compared to the observed difference in spectral resolution in the data between the M and N regions noted above. The Dy M1 binding energy used for energy calibration of the measured spectrum is the same as that used in the calculation. The value of the broadening parameter was selected by eye to provide a reasonable match between calculation and data. Besides these three factors (energy calibration, vertical scale, and broadening), the calculation has no other adjustable parameters and directly uses the formulas from Ref. [18] with the same numerical values. Of particular note is that the weighting factor for each peak comes from theoretical nuclear-atomic physics calculations (e.g. wave function overlap integrals, inclusion of Dy states with up to three holes).

The measured and calculated spectra are very similar, showing many of the same key features, in the same places and with about the same size, though differing in details. The energy of the M1 peak and the height of the N1 peak provide no new information because they are used for energy calibration and vertical scaling respectively. The height of the calculated M1 peak is about 50% higher than the measured spectrum, though this might be compensated by the undershoot of the calculation from about 1900-2000 eV. The high-energy side of the M1 peak appears to match well. For the M2 peak, the energy location and local spectral shapes of the calculated and measured peaks are reasonably consistent given the limited statistics (fewer than 50 counts in any 5 eV bin near the M2 peak). The energy locations of the calculated and measured N1 peaks are reasonably consistent. The lower-energy side of the N1 peaks follow each other in calculation and data, with a discrepancy emerging on the higher-energy side. There is a shoulder observed at about 450-500 eV in the data that is not seen in the calculation.

The N2 peak occurs at about the same energy location and size in the calculation and the data. Other potential features interpretable as similar or different between the calculated and measured spectra all occur in energy ranges with fewer than ten counts in any 5 eV bin. Therefore, these potential similarities and differences should be viewed skeptically, and more data will be needed for a useful assessment. Nonetheless, we are intrigued by the future possibility of observing the predicted 2-hole or 3-hole peaks, e.g. the M1+N4/5 (3s+4d) state calculated at 2201.8 eV and the N1+N4/5 (4s+4d) state calculated at 569 eV [18].

## 6 Summary

We have demonstrated a complete process for production, isolation, and purification of $^{163}$Ho, and its incorporation into transition-edge sensor microcalorimeters developed for this purpose. Our measured $^{163}$Ho electron-capture spectrum shows peaks that correspond to the M1, M2, N1, and N2 states of the $^{163}$Dy daughter atom. We have shown that measured and calculated spectra are very similar, with many of the same key features consistent in energy and intensity, though details differ.

**Acknowledgements**
We gratefully acknowledge support of the Laboratory-Directed R&D Program of Los Alamos National Laboratory; the U. S. Department of Energy, Office of Science, Nuclear Physics, Isotope Program; and the Center for Integrated Nanotechnologies, an Office of Science User Facility. We thank M. Caro and J. K. Baldwin for assistance in fabrication of nanoporous gold, and A. Faessler and A. De Rujula for discussion of ECS theory.